\documentclass[12pt]{article}

\usepackage[margin=1in]{geometry}
\usepackage{graphicx}
\usepackage{enumerate}
\usepackage{url}
\usepackage[super,sort&compress, comma]{natbib}
\usepackage{soul}
\usepackage{mathptmx}
\usepackage{macros}
\usepackage[section]{placeins}
\usepackage{subfigure}
\usepackage{color}
\usepackage{makecell}

\newcommand{\Ar}[0]{{A^R_{\ell,b}}}
\newcommand{\Ad}[0]{{A^D_{\ell,b}}}
\newcommand{\lar}[0]{{\ell, \Ar}} 
\newcommand{\lad}[0]{{\ell, \Ad}}

\newcommand{\ifr}[0]{\text{IFR}}
\newcommand{\Bin}{\text{Binomial}}
\newcommand{\sens}{\text{sens}}
\newcommand{\spec}{\text{spec}}

\newcommand{\blind}{0}

\begin{document}

\def\spacingset#1{\renewcommand{\baselinestretch}%
{#1}\small\normalsize} \spacingset{1}


\date{}

\if0\blind
{
    \begin{center}
    {\LARGE\bf A hierarchical Bayesian model for estimating age-specific COVID-19 infection fatality rates in developing countries}
    \vspace{2em}
    
    { Sierra Pugh\hspace{.2cm}\\
    Department of Statistics, Colorado State University\\
    and \\

    Andrew T. Levin \\
    Department of Economics, Dartmouth College \\
    National Bureau for Economic Research \\
    Centre for Economic Policy Research \\
    and \\
    Gideon Meyerowitz-Katz \\
    School of Health and Society, University of Wollongong \\
    Western Sydney Local Health District \\
    and \\
    Satej Soman \\
    School of Information, UC Berkeley \\
    and \\
    Nana Owusu-Boaitey \\
    Case Western Reserve University School of Medicine \\
    and \\
    Anthony B. Zwi \\
    School of Social Sciences, University of New South Wales \\
    and \\
    Anup Malani \\
    National Bureau for Economic Research \\
    Law School, University of Chicago \\
     and \\
     Ander Wilson \\
     Department of Statistics, Colorado State University \\
     and \\
        Bailey K. Fosdick \\
    Department of Biostatistics and Informatics, Colorado School of Public Health \\
   }
    \end{center}
} \fi

\if1\blind
{
  \bigskip
  \bigskip
  \bigskip
  \begin{center}
    {\LARGE\bf A hierarchical Bayesian model for estimating age-specific COVID-19 infection fatality rates in developing countries}
\end{center}
  \medskip
} \fi

\bigskip
\begin{abstract}
The COVID-19 infection fatality rate (IFR) is the proportion of individuals infected with SARS-CoV-2 who subsequently die. As COVID-19 disproportionately affects older individuals, age-specific IFR estimates are imperative to facilitate comparisons of the impact of COVID-19 between locations and prioritize distribution of scare resources.
However, there lacks a coherent method to synthesize available data to create estimates of IFR and seroprevalence that vary continuously with age and adequately reflect uncertainties inherent in the underlying data. In this paper we introduce a novel Bayesian hierarchical model to estimate IFR as a continuous function of age that acknowledges heterogeneity in population age structure across locations and accounts for uncertainty in the estimates due to seroprevalence sampling variability and the imperfect serology test assays. Our approach simultaneously models test assay characteristic, serology, and death data, where the serology and death data are often available only for binned age groups. Information is shared across locations through hierarchical modeling to improve estimation of the parameters with limited data. Modeling data from 26 developing country locations during the first year of the COVID-19 pandemic, we found seroprevalence did not change dramatically with age, and the IFR at age 60 was above the high-income country benchmark for most locations. 
\end{abstract}

\vfill

\newpage

\spacingset{1.8}

\section{Introduction}

The 2019 coronavirus disease (COVID-19) pandemic has had devastating impacts worldwide, with at least 6.59 million confirmed COVID-19 deaths as of November 2022 \citep{jha2022counting, deaths}. However, comparing the burden across locations with varying age distributions is specifically difficult because COVID-19 infections and fatalities vary substantially by age \citep{eje, pezzullo2023age, starke2021isolated}. Estimating age-specific metrics of the COVID-19 burden is, therefore, necessary to make meaningful comparisons across locations and best allocate scarce resources with age-specific policies.   
Between location comparisons are further complicated by small sample sizes and uncertainty about testing characteristics of newly developed COVID-19 tests, such as sensitivity and specificity, both of which result in additional uncertainty about location-specific estimates.
Policy makers turned to these studies to make real-time decisions, and therefore, it was essential to quantify and communicate uncertainty in the estimates. In this study, we focus on estimating age-specific metrics of the COVID-19 burden, along with corresponding uncertainty bounds, and comparing the burden across developing country locations with different age distributions.

There are several possible measures of disease burden. In this paper, we focus on estimating prevalence and the infection fatality rate (IFR)--the proportion of individuals infected with severe acute respiratory syndrome coronavirus 2 (SARS-CoV-2) that subsequently die from the disease. Compared to the more commonly used and easier to estimate case fatality rate (CFR)--defined as the ratio of the number of deaths to the number of reported cases--IFR depends on a measure of the number of prior infections, including unreported infections. A key benefit of IFR is it not biased by preferential testing and reporting. 

Early in the COVID-19 pandemic, tests for active infection of COVID-19 were limited and essentially unavailable in many areas. Thus, reported cases were likely a substantial under count of infections \citep{national2020evaluating} and researchers instead turned to serology studies to estimate the number of infections. In a serology study, a sample of the population is tested with an antibody test to identify those who have had a COVID-19 infection and seroconverted, i.e., built antibodies against the virus. The proportion of individuals with SARS-CoV-2 antibodies--the seroprevalence--is then used as a proxy for the proportion of the population that has been infected. Antibody tests used early in the pandemic were developed quickly using limited validation data to estimate test characteristics such as specificity and sensitivity. As a result, estimates of prevalence from serology studies were plagued by uncertainty due to often small samples from the population, inaccuracies from the antibody tests, and further general uncertainty in the antibody test's characteristics. Accounting for these avenues of uncertainty was paramount in reporting uncertainty in community prevalence \citep{gelman2020bayesian}.   

Discrepancies between the effects of COVID-19 for the young and old was quickly recognized.   
Levin et al\citep{eje} estimated that variation in age distributions between countries explained approximately 90\% of the variation in population-level IFRs. Specifically, they found the IFR for high-income countries to be about 0.01\% at age 25 and about 15\% at age 85. Thus, age-specific IFR estimates are required to make meaningful comparisons of the disease burden across locations and inform age-based policies that mitigate disease transmission \citep{malani2022vaccine}. This relies upon information on age-specific infections and age-specific fatalities. 
Most studies of COVID-19 IFR reported estimates for coarse age bins \citep[e.g.,][]{perez, odriscoll, pezzullo2023age}. A limited number of studies have reported estimates of COVID-19 IFR as a continuous function of age \citep{eje, ours, lancet}, but all of these studies rely on multi-step modeling approaches and lack location-specific curves. 

The majority of studies quantifying the burden COVID-19 have focused on high income countries, with large serology studies and highly granular death data.  
The data availability and quality in developing counties posses additional challenges for estimating age-specific rates and between country comparison. In developing countries, serology studies typically had smaller sample sizes and reported data on age inconsistently, with some locations providing individual data and other reporting only coarse age bins. 
As a result, existing methods that rely on detailed infection and death data are not suitable. 

In this study, we focus on estimating infection and fatality rates for 26 developing country locations, where serology studies were reported for coarse age bins with often low sample sizes. 
We propose a Bayesian hierarchical modeling approach to jointly estimate continuous age-specific and location-specific seroprevalence and IFR curves that account for the many aforementioned challenges: the many sources of uncertainty, binned nature of seroprevalence and death data, and the inconsistent and coarse data resolution at which the developing country data are reported. By jointly modeling seroprevalence, antibody test assay characteristics, and the IFR, we fully propagate uncertainty to all model estimates. Our hierarchical modeling framework shares information across age bins within a location, as well as across locations to improve estimation in cases of limited data.  While prior works have addressed some aspects of this problem, we present a holistic solution that addresses all objectives simultaneously. Using our proposed model, we estimate continuous seroprevalence curves for all locations, which allow for age-specific comparisons between our developing country study locations and high-income country estimates.

\section{Data}\label{data}

In this study we revisit the serology and death data from developing countries described in Levin et al\cite{ours}. The rich data set contains seroprevalence study data for a total of 107 locations from 44 countries. Of the 107 locations,
63 had corresponding COVID-19 fatality data. 
Lab validation data was assimilated from antibody test manufacturers and the population age distribution was gathered for each location. The data sources vary in terms of information provided, age aggregation, antibody test and test characteristics, and quality. 
 
\begin{figure}[h!]
    \centering
    \includegraphics[width=\textwidth, clip, trim=.45in 0in .05in .35in]{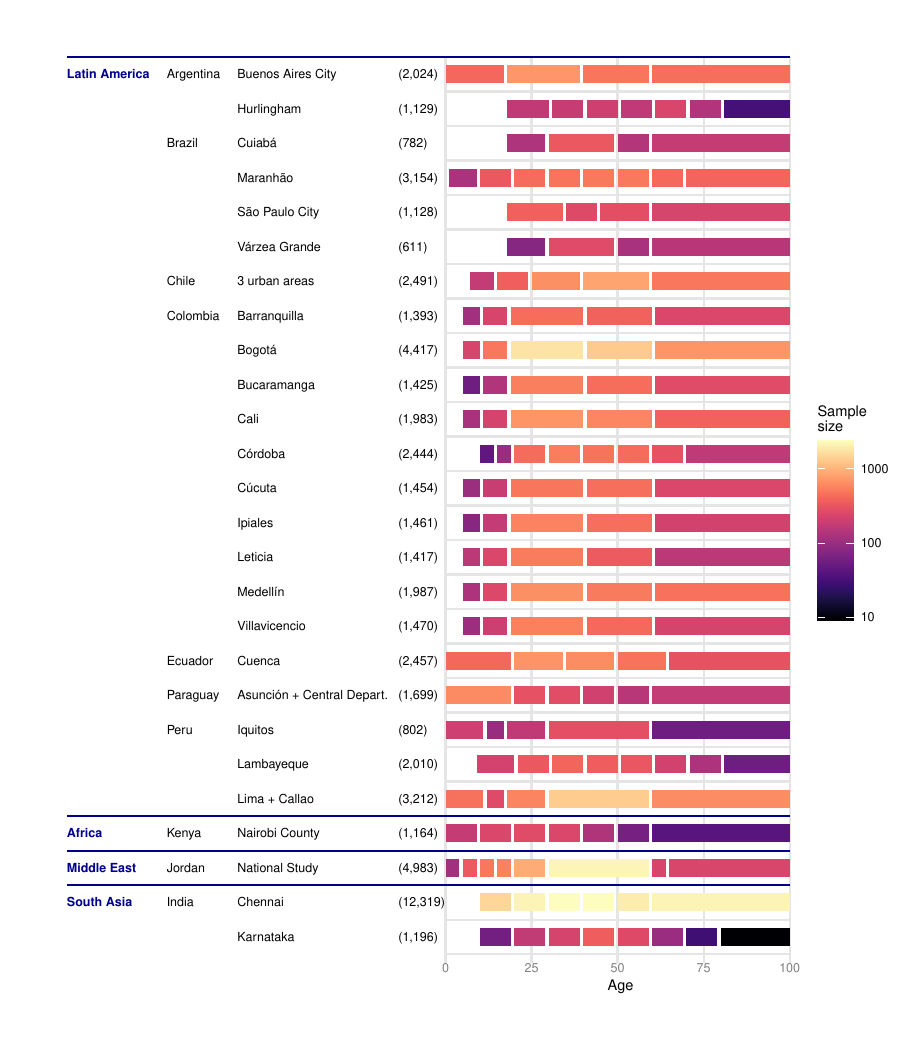}
    \caption{Number of participants in serology study for each age bin and location ($n_\lar$). Total sample size shown in parentheses.}
    \label{fig:serobins}
\end{figure}

Because our aim is to estimate seroprevalence and IFR as flexible, continuous functions of age, we restricted our analysis to studies with age-specific seroprevalence and death data. Making inference about the shape of a curve from only a few data points is extremely challenging, and prior analysis found IFR showed heterogeneity across locations that was not well explained by observed covariates, such as healthcare capacity or GDP, limiting the potential of borrowing information in a data-informed manner \citep{ours}. 
Therefore, we restrict our attention to studies with at least four serology age bins and at least five death age bins to allow inference on the structure of IFR with age, resulting in a total of 26 study locations for this analysis. 
Twenty-two locations are from Latin America, two are from South Asia, and there is one study each from Africa and the Middle East (see Figure \ref{fig:serobins}). Each study focused on a specific region within the country, with the exception of Jordan, which is a national study. For example, there are nine studies within Colombia, each confined to a different, non-overlapping region. 
The data that support the findings of this study are openly available in Zenodo at http://doi.org/ 10.5281/zenodo.8048531 \cite{ourdata}.

\subsection{Serology data}
As described in Levin et al\cite{ours}, we collected and evaluated serology studies from government reports, published papers, and preprints. Studies using convenience samples like blood donors, volunteers, or residual sera were excluded, as well as those with imbalanced gender ratios, inaccurate test assays, insufficient reported data, and studies occurring during accelerating outbreaks. All studies were concluded before March 2021--before vaccines were readily available in developing countries. Therefore, a positive antibody test due to a COVID-19 vaccine, rather than a prior infection, is extremely unlikely in the data. 

\begin{figure}[bth]
  \centering
  \subfigure[]{%
    \begin{tabular}[b]{crr}
    \hline
        \makecell{Age bin \\ $\left(\Ar \right)$} & \makecell{Sample size \\ $\left(n_{\lar}\right)$} & \makecell{Number positive  \\ $\left(R^\star_{\lar}\right)$} \\ \hline
9-20 & 220 & 59 \\ 
  21-30 & 329 & 102 \\ 
  31-40 & 394 & 139 \\ 
  41-50 & 353 & 109 \\ 
  51-60 & 315 & 94 \\ 
  61-70 & 215 & 54 \\ 
  71-80 & 129 & 26 \\ 
  81+ & 55 & 12 \\ 
   \hline
  \vspace{2.5em}
    \end{tabular} 
    \label{}
  }
  \subfigure[]{\raisebox{1.5em}{\includegraphics[width=.45\textwidth,
  clip, trim=0in 0in 0in .05in]{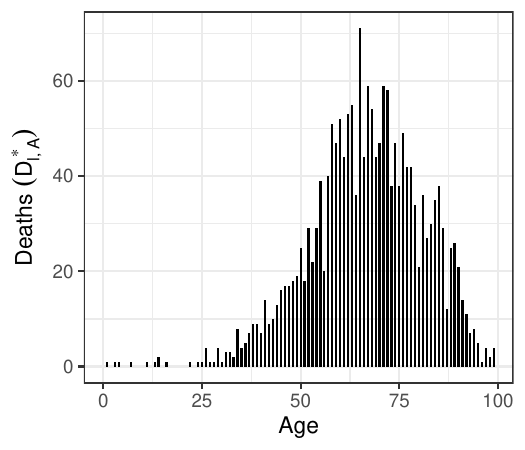}} }
  \qquad
  \caption{\label{tab:lambayeque}(a) Seroprevalence data collected between 6/24/2020 7/10/2020 and (b) corresponding cumulative death data ($D^\star_\lad$) for Lambayeque, Peru. }
\end{figure}

For each location, researchers reported results for between four and eight age bins. The total number of participants tested and the number of participants that tested positive were then recorded for each bin. Figure \ref{fig:serobins} shows the age bins for each study location, as well as the number of participants (i.e., sample size) in each bin. Most locations have wider age bins and often smaller sample sizes for the oldest individuals as seen in, for example, Hurlingham, Argentina and Lambayeque, Peru. Karnataka, India had the smallest average bin sample size with between 9 and 353 samples for each age bin, whereas Chennai, India had the largest average bin sample size with between 1473 and 2353 samples for each bin. Panel (a) of Figure \ref{tab:lambayeque} shows an example of the seroprevalence data for Lambayeque, Peru. 

\FloatBarrier

\subsection{Test characteristic data}
\begin{figure}[tb]
    \centering
    \includegraphics[scale=.75]{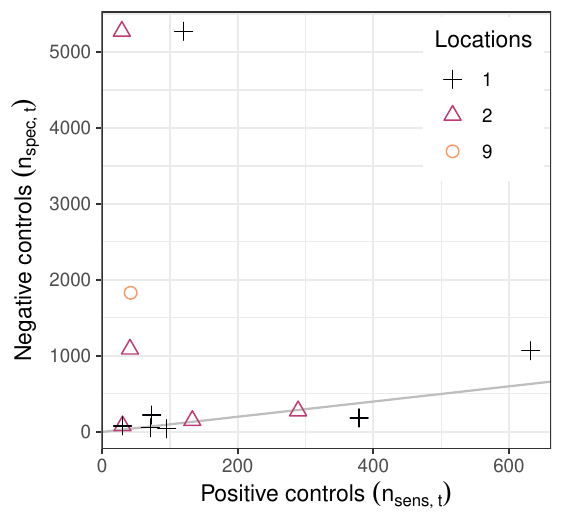}
    \caption{Number of positive controls ($n_{\text{sens},t}$) and number of negative controls ($n_{\text{spec},t}$) for each test assay. The color and shape of the point indicate the number of locations using the assay, either 1, 2, or 9. The line $x=y$ is shown in grey.}
    \label{fig:controls}
\end{figure}

For each test used in a seroprevalence study, we directly model the test assay lab validation data, namely the number of positive/negative controls tested in the validation study and the number that test positive in the validation study, to estimate the test sensitivity and specificity. In total, thirteen tests were used in our sample of seroprevalence studies, with six of the test assays used in multiple studies (see Figure \ref{fig:controls}). Between 29 and 632 positive controls and between 42 and 5272 negative controls were used to validate each test assay. For instance, the Coretest COVID-19 IgM/IgG Antibody Test used in Lambayeque, Peru was developed using 73 positive controls, 58 of which tested positive, giving a crude sensitivity estimate of 0.795, and 222 negative controls, which all correctly tested negative, giving a crude specificity estimate of 1. 

\subsection{Death data}

We modeled aggregate data on the number of confirmed and suspected COVID-19 fatalities associated with each serology study location using individual case data, when available, and public health reports otherwise. 
To account for the delay in time between infection and fatality, we collected the cumulative number of deaths up to fourteen days after the midpoint date of the associated seroprevalence study when using individual case data and the cumulative number of deaths up to twenty-eight days after the midpoint date of the study when using public health reports due to the lag in death reporting. 

Twenty study locations had individual case data available, from which we can determine the number of deaths for each one-year age bin (e.g., $[0,1)$, $[1,2)$, etc.). For the remaining six locations, death data was available for between six and eighteen age bins. Lambayeque, Peru, for example, has individual case data summarized in one-year age bins (Figure \ref{tab:lambayeque} panel (b)). Fifteen ages had zero deaths, with the most deaths, namely 71, occurring at age 65. 

\subsection{Age distribution data}

The population distribution data, collected from census data and websites such as worldometers.info, worldpopulationreview.com, and populationpyramid.net, contained the number of people in each study area at a resolution of between seven to twenty age bins. These distributions were further refined to a 5-year age bin resolution using national age distribution data (see Section 1 of the Supplementary Materials).  
Acknowledging that age is a fundamentally continuous measure, our model operates on a density function of age. For application to this developing country data set, the 5-year binned age distribution data was further converted to a continuous function of age by  
fitting a locally weighted linear regression line. The resulting smooth function for the age distribution was then scaled to integrate to one to create a density. While a simpler approach would have been to assume the population is uniform within an age bin, this assumption is clearly violated for older age groups where the risk of death is largest and most variable. These facts motivated our more complex approach, which is detailed further in Section 1 of the Supplementary Materials.

\FloatBarrier
  
\section{Methods}\label{model}

\begin{figure}
    \centering
\includegraphics[width=.9\textwidth, clip, trim=.6in 1.2in 5.2in 1.2in]{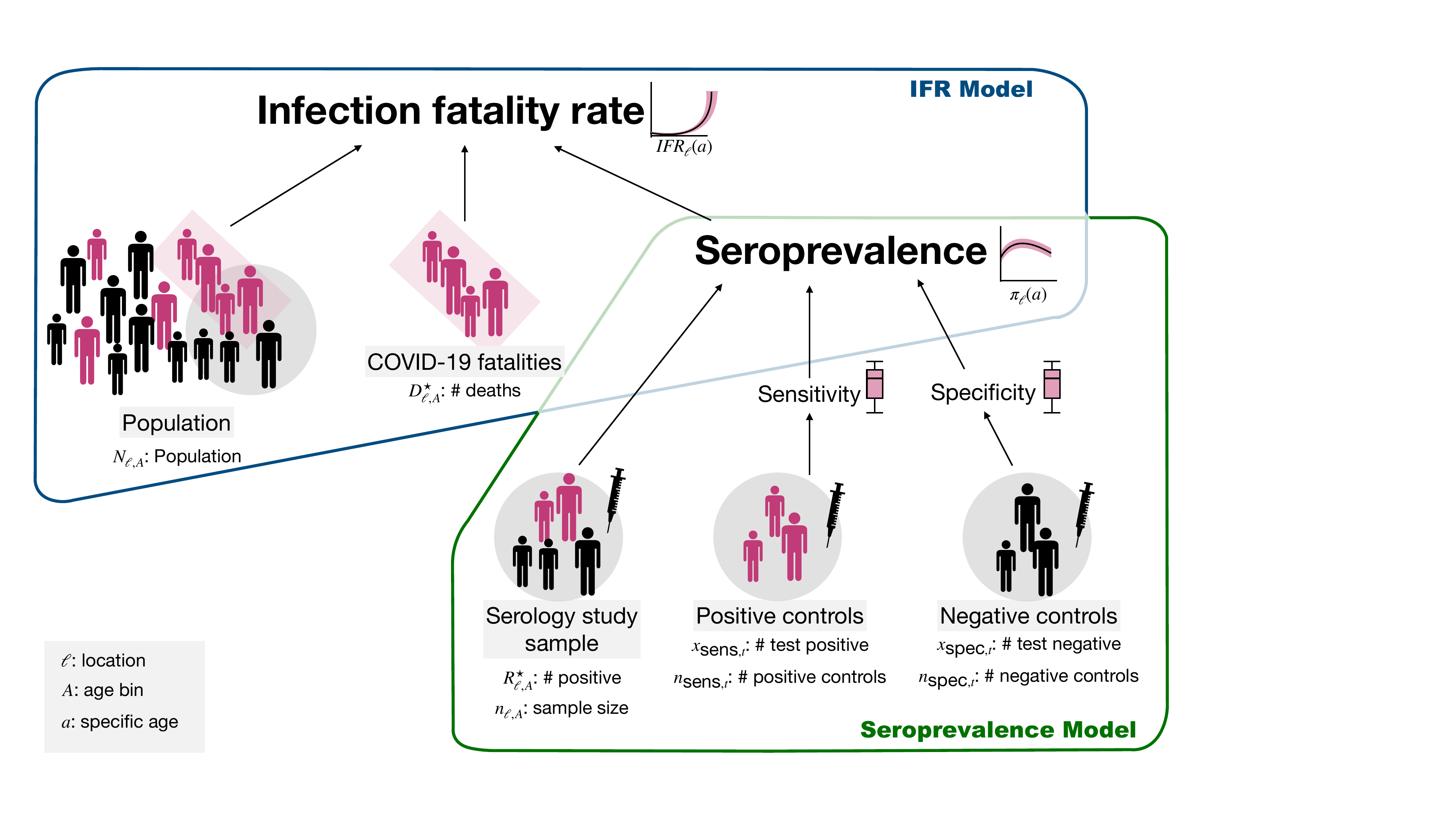}
    \caption{The relationship between our data sources and the seroprevalence and IFR functions of interest. Note, this represents a single location and a single age bin within this location.}
    \label{fig:diagram}
\end{figure}

We propose a joint model for seroprevalence and IFR using a Bayesian hierarchical framework. Our approach includes two submodels to combine the many data sources shown in Figure \ref{fig:diagram}. First, we use a logistic regression model to model the serology study data as a function of age. To account for uncertainty in the test characteristics, we use binomial models for the positive and negative controls to model sensitivity and specificity. Second, we model the fatality data using a Poisson regression as a function of age and estimated age-specific seroprevalence. Together, these submodels allow for location- and age-specific seroprevalence and IFR estimates that fully account for the uncertainty of each data source.

\subsection{Modeling seroprevalence}\label{sero}

We seek to model seroprevalence as a continuous function of age from the serology study data that is only reported for age bins. We assume the observed data is informative about the (population weighted) average seroprevalence within the respective age bin. Specifically, we define this average seroprevalence as the integral of the continuous seroprevalence function with respect to the population's age distribution over the bin. In the rest of this section, we describe our model for the continuous seroprevalence function and the test validation data.

\subsubsection{Modeling observed serology data}

Let $\pi_\ell(a)$ represent the unknown seroprevalence rate for location $\ell$ and age $a$, one of our primary parameters of interest. We define $b\in \{1, \dots, B_\ell^R\}$ as indices for the serology age bins at location $\ell$. Let $\Ar$ denote an age bin of the serology study at location $\ell$, $R^\star_\lar$ denote the number of individuals who tested positive in age bin $\Ar$ in study $\ell$, and $n_\lar$ denote the number of individuals tested in age bin $\Ar$ at location $\ell$.  

We model the number of people who test positive as coming from a binomial distribution 
\begin{align}\label{seromod}
    R^\star_\lar &\sim \Bin(n_\lar, p_\lar),
\end{align}
where the average test positivity in that age bin is represented by $p_\lar$. The test positivity represents the proportion of individuals' tests we expect to be positive, while the seroprevalence represents the proportion of individuals with COVID-19 antibodies (i.e., are seropositive). 
Utilizing a binomial distribution assumes the tests are independent of each other, which aligns with the study inclusion criteria of Levin et al\cite{ours} that required samples to be representative.

We model the average positivity for age bin $\Ar$ at location $\ell$, $p_\lar$, as the integral of a continuous age-specific positivity function, $p_\ell(a)$, which is integrated with respect to the location $\ell$'s population age distribution.  Mathematically this can be expressed 
\begin{align}\label{eq:pavg}
    p_\lar &= \int_{\Ar} p_\ell(a) \frac{f_\ell(a)}{\int_{\Ar} f_\ell(x) dx} da,
\end{align}
where $f_\ell(a)$ denotes the population age density at location $\ell$ evaluated at age $a$. The age distribution is normalized by the total population in age bin $\Ar$, so $p_\lar$ is a weighted average positivity rate, weighted by the relative population within the interval. Details for defining $f_\ell(a)$ are given in Section 1 of the Supplementary Materials. 

Since serology tests are not perfectly accurate, false positive results and false negative results are expected, and the frequency of these depends on the true seroprevalence and characteristics of the test \citep{gelman2020bayesian}. For the test assay $t_\ell$ used at location $\ell$, the relationship between the test positivity rate and true seroprevalence is
\begin{align}\label{eq:rg}
    p_\ell(a) &= \pi_\ell(a) \sens_{t_\ell} + (1-\pi_\ell(a)) (1-\spec_{t_\ell}),
\end{align}
where $\sens_{t_\ell}$ and $\spec_{t_\ell}$ represent the sensitivity and specificity of the test, respectively. The first term in $p_\ell(a)$ equals the proportion of the population correctly identified as seropositive and the second term is the proportion of the population incorrectly identified as seropositive.

\subsubsection{Seroprevalence as a continuous function of age}

 We model seroprevalence for location $\ell$ and age $a$, $\pi_\ell(a)$, as a linear function of covariates on the logit scale:
\begin{align}\label{eq:regsero}
    \text{logit} \left(\pi_\ell(a) \right) &= \gamma_{\ell,0} + \vec{z}'_{\ell,a}\vec\gamma_\ell,
\end{align}
where $\gamma_{\ell,0}$ is an intercept, $\vec{z}_{\ell,a}$ is a $p$-dimensional vector of covariates, and $\vec\gamma_\ell=(\gamma_{\ell,1}, \dots, \gamma_{\ell,p})'$ is a $p$-dimensional vector of coefficients. Because seroprevalence is a proportion, we use the logit link to constrain $\pi_\ell(a)$ to be between zero and one.  
We are specifically interested in seroprevalence as a function of age and, therefore, specify $\vec{z}'_{\ell,a}$ to be a natural spline of age. Yet, our framework is general and can accommodate any covariates of interests. 
For example, population density in a location or an indicator for whether the location underwent government delegated stay-at-home orders could be added as covariates. These may be informative predictors if seroprevalence is higher in denser populations or mobility restrictions prevented transmission leading to lower seroprevalence. In our application, we lacked rich covariate information on all locations, so $\vec{z}'_{\ell,a}$ was specified to be a natural spline of age (details given in Section \ref{appCovariates}).

\subsubsection{Modeling test validation data}

At the onset of the pandemic, serological test assays were developed quickly with limited controls. Therefore, we consider sensitivity and specificity to be unknown parameters and explicitly model the lab validation data as done in Gelman and Carpenter\citep{gelman2020bayesian}, Stringhini et al\citep{stringhini2020seroprevalence}, and Larremore et al\citep{larremore2022optimizing}. 
Let $n_{\sens,t}$ denote the number of positive controls and $n_{\spec,t}$ denote the number of negative controls for test assay $t$. Further, let $x_{\sens,t}$ denote the number of positive controls that tested positive and $x_{\spec,t}$ denote the number of negative controls that tested negative. Naive point estimates of the test sensitivity and specificity are then $x_{\sens,t}/n_{\sens,t}$ and $x_{\spec,t}/n_{\spec,t}$, respectively. Rather than taking the test characteristics as known, 
we estimate $\spec_t$ and $\sens_t$ along with their uncertainty from the lab validation data. The model is
\begin{align}
    x_{\sens,t} &\sim \Bin(n_{\sens,t}, \sens_t), \\
    x_{\spec,t} &\sim \Bin(n_{\spec,t}, \spec_t). 
\end{align}
We assume each control sample tested is a Bernoulli trial, where positive controls test positive with probability equal to the sensitivity of the assay, and negative controls test negative with probability equal to the specificity of the assay. 

When a test is used at multiple study locations, the sensitivity and specificity are assumed to be the same in each study. However, we do not assume any relationship between the sensitivity and specificity across tests. 
One reason for this is that antibody tests are designed to target different types of antibodies, such as IgG versus IgM, as well as rely on binding different regions of SARS-CoV-2 \citep{jacofsky2020understanding}. Therefore, while we pool information across locations that use the same test, we do not pool information about the sensitivity and specificity across tests.

Seroprevalence, sensitivity, and specificity jointly determine the positivity rate. 
With small sample sizes, sensitivity and specificity are weakly identified. We have prior information on the target test characteristics for antibody tests, so we use informative priors to improve identifiability. 
The priors selected for the developing countries data set specifically are discussed in Section \ref{priors}.

\subsection{Modeling IFR}\label{ifr}

The model for IFR is similar to the model for seroprevalence as we model IFR as a function of age and other covariates. We link the continuous age-specific IFR parameter to the data by calculating the average IFR at the observed age bins via integration.

\subsubsection{Modeling observed death counts}

Let $\Ad$ denote an age group for which deaths caused by COVID-19 are recorded, $b \in \{1, \dots, B^D_\ell\}$. We model the $D^\star_\lad$ deaths at location $\ell$ and age bin $\Ad$ as following a Poisson distribution: 
\begin{align}\label{deathmod}
    D^\star_\lad &\sim \text{Poisson}(N_\lad \Lambda_\lad),
\end{align}
where $N_\lad$ denotes the population at location $\ell$ in age group $\Ad$, and $\Lambda_\lad$ represents the proportion of individuals in location $\ell$ and age group $\Ad$ expected to die from the disease. Thus, the product $N_\lad \Lambda_\lad$ represents the number of individuals in that location and age bin expected to die from the disease. Similar to O'Driscoll et al\citep{odriscoll}, we chose to model deaths with a Poisson distribution rather than a binomial distribution because COVID-19 deaths are a relatively rare event. Note, deaths are modeled as a census of the population, assuming deaths are accurately reported.

The death rate for age $a$ is given by $\pi_\ell(a) \times \ifr_\ell(a)$: the probability of infection times the probability of death given infection at age $a$. Similar to the average positivity in \eqref{eq:pavg}, we define the population weighted average death rate for age bin $\Ad$, $\Lambda_\lad$, as 
\begin{align}\label{eq:lambdaAvg}
    \Lambda_\lad &= \int_\Ad \pi_\ell(a) \times \ifr_\ell(a) \frac{f_\ell(a)}{\int_\Ad f_\ell(x) dx} da.
\end{align}
There are infinitely many possible combinations of seroprevalence and IFR that can result in the same $\Lambda_\lad$, so we recommend an informative prior for at least one of these when the serology data is limited. In our analysis, we apply informative priors to the parameters in the serology model as described in Section \ref{priors}.

\subsubsection{IFR as a continuous function of age}

Let $\ifr_\ell(a)$ be the IFR at location $\ell$ and age $a$. We define $\log(\ifr_\ell(a))$ as a function of a natural spline of age, $\vec{x}_{\ell,a} \in \mathbb{R}^q$ and associated location-specific coefficients $\vec\beta_\ell =(\beta_{\ell,1}, \dots, \beta_{\ell,q})'$
\begin{align}\label{eq:regifr}
    \log\left(\ifr_\ell(a)\right) &= \beta_{\ell,0} + \vec{x}'_{\ell,a}\vec\beta_\ell,
\end{align}
where $\beta_{\ell,0}$ represents the study location-specific intercept. This framework is general and can incorporate additional covariates. 

Since we expect the IFR to be generally similar across study locations, we use hierarchical priors to pool information about $\vec\beta_\ell$ across locations. 
We further expect overall IFR levels to be more similar for locations within the same country compared to study locations in different countries, so we model the IFR intercept with a common country effect. This is particularly relevant for developing countries as the death registration systems can differ dramatically between countries  
\citep{karanikolos2020comparable}.
Thus, we define priors on the coefficients as
\begin{align}
    \beta_{\ell,0} &\sim \N(\beta_{\text{global},0} + \beta_{\text{country},c_\ell}, \sigma^2_0), \\
    \beta_{\ell,i} &\sim \N(\beta_{\text{global},i}, \sigma^2_i), \qquad \text{ for i } \in \{1, \dots, q\}.
\end{align}
The intercept ($\beta_{\ell,0}$) is informed by an overall global intercept parameter ($\beta_{\text{global},0}$) as well as a country specific intercept ($\beta_{\text{country},c_\ell}$). The other covariates, which control the shape of the IFR function, pool information at the global level using $\beta_{\text{global},i}$, but not at the country level. 

\subsection{Covariate and prior distribution selection} \label{applyDevelopingCountries}

The methods have been discussed rather generally up until this point as they are applicable to any disease setting. However, we now focus on modeling choices made to account for the specific nuances of our application: COVID-19 in developing countries from June 2020 to March 2021. Specifically, we discuss the covariates and the prior distributions in this section. 

\subsubsection{Covariates}\label{appCovariates}

For modeling IFR as a function of age, we used natural cubic splines with boundary knots at 0 and 80, and internal knots at 10 and 60 as the covariates, $\vec{x}_{\ell,a}$. Thus, log IFR is modeled as cubic functions of age between 0 and 80 and is constrained to be log-linear above age 80. The knots were selected based on expert opinion and prior literature \citep{anup, lancet}. 
Similarly, we used natural cubic splines for the seroprevalence covariates, $\vec{z}_{\ell,a}$, with an internal knot at 60 and the boundary knots at 10 and 80. We decreased the number of internal knots in the natural spline for the serology model because each location in our data set contains only three to eight age bins, with over half the observations having five bins.

\subsubsection{Prior specification} \label{priors}

Estimation of our model is straightforward in a Bayesian context. 
Similar to Gelman and Carpenter\citep{gelman2020bayesian}, we set independent, informative beta priors for the sensitivity and specificity of each of the $T=13$ tests:
\begin{align}
    \sens_t &\sim \text{Beta}(10,1), \qquad \text{for } t \in \{1, \dots, 13\}, \\
    \spec_t &\sim \text{Beta}(50,1), \qquad \text{for } t \in \{1, \dots, 13\}.
\end{align}
Seroprevalence assays are generally designed to ensure a high specificity. For example, the Centers for Disease Control and Prevention recommended a specificity of 0.995 for COVID-19 antibody tests \citep{cdcguidelines}. The prior specified for specificity has a 0.9 probability the specificity is greater than 0.95. Because tests are designed to prioritize specificity, the sensitivity is typically more variable, reflected by our more dispersed prior with 0.9 probability the sensitivity is greater than 0.79. 

We utilize informative priors on the coefficients for the seroprevalence function at each location based on prior studies. Specifically, we specify independent normal distributions centering the non-intercept coefficients about zero:
\begin{align}
    \gamma_{\ell,0} &\sim \N(-1, 1.5), \\
    \gamma_{\ell,j} &\sim \N(0, 0.05), \qquad \text{for } j \in \{1, 2\}.
\end{align}
Here $\gamma_{\ell,0}$ denotes the intercept at location $\ell$ and $\gamma_{\ell,j}$ for $j>0$ are the coefficients associated with the natural spline, controlling the shape of the seroprevalence function. Because seroprevalence is modeled as a logit-linear function of the covariates, the prior on the intercept is a right skewed distribution with a median of about 0.27 and a 95$^\text{th}$ quantile of about 0.81. The more informative priors on the spline coefficients ($j>0$) have a prior mean of zero, corresponding to a seroprevalence function that does not vary with age, which has been noted in the literature.  Levin et al\citep{ours} found the ratio of seroprevalence for those age 60+ compared to 40-59 was not significantly different from one in most locations. Esteve et al\citep{esteve2020national} also found countries where multigenerational households are common, such as developing countries \citep{ruggles2008intergenerational}, may not be able to shield their older population from COVID-19.  
 
We use weakly-informative priors for the age spline coefficient parameters associated with IFR:
\begin{equation}
\begin{aligned}
    \beta_{\text{global},i}  &\sim \N(0,5), \qquad &\text{for } i \in \{0, \dots, 3 \},\\
    \beta_{\text{country},c_\ell} &\sim \N(0, \sigma_\text{country}), \\
    \sigma_i &\sim \text{half-normal}(0,2), \qquad &\text{for } i \in \{0, \dots, 3\}, \\
    \sigma_\text{country} &\sim \text{half-normal}(0,2).
\end{aligned}
\end{equation}
Note, IFR is modeled on the log scale, so these priors are relatively non-informative. For example, with a prior probability of 0.8, the global intercept, $\beta_{\text{global},0}$ is between -6.4 and 6.4, corresponding to a multiplicative effect between 0.002 and 606.5. 

\subsection{Estimation and Inference} \label{Estimation}

Inference was based on the joint posterior distribution of the IFR parameters $(\{\beta_{\ell,i}\}, \{\beta_{\text{global},i}\}, \\ \{\beta_{\text{country},c}\}, \{\sigma_i\}, \sigma_\text{country})$ and serology parameters $(\{\gamma_{\ell,j}\}, \{\sens_t\}, \{\spec_t\})$ given the death, serology, and antibody test validation data $(  \{n_\lar\}, \{R^\star_\lar\}, \{D^\star_\lad\},  \{x_{\sens, t}\}, \{x_{\spec,t}\}, \{n_{\sens,t}\}, \\ \{n_{\spec,t}\})$. The joint posterior is not available in closed form, so we obtain posterior draws through a Markov chain Monte Carlo (MCMC) algorithm. We approximated the integrals in \eqref{eq:pavg} and \eqref{eq:lambdaAvg}, which calculate the average positivity and death rates for the age bins, using trapezoidal Reimann sums scaling the mesh appropriately to increase computation speed, while maintaining sufficient accuracy. The MCMC algorithm was implemented in version 2.21.5 of RStan \citep{stan}.

We use the $\gamma_{\ell,j}$ and $\beta_{\ell,j}$ posterior draws to calculate the location-specific IFR and seroprevalence functions of age as in \eqref{eq:regifr} and \eqref{eq:regsero}. These allow for age-specific comparisons, rather than age bin level comparisons. We summarized the parameters of interest in terms of posterior means and 95\% credible intervals. Note, we do not interpret $\beta_{\text{global},i}$ as a ``global average" in the geographic sense since the study locations are not a representative sample of the world as a whole. Similarly, $\beta_{\text{country},c_\ell}$ is not interpreted as the adjustment for an entire country since the study locations are not a representative sample of locations within a country.

\section{Analysis of the developing countries data} \label{application}

To sample from the posterior distribution, we ran three chains with a burn-in of 2500 iterations and the subsequent 3000 samples were retained from each chain.  Code is available at \if0\blind
{\url{https://github.com/pughs/covid-ifr}}\fi \if1\blind {\url{url}}\fi. Convergence was assessed via traceplots, effective sample size, and $\hat{R}$ \citep{gelman1995bayesian}. Each parameter had an effective sample size greater than 1000 and $\hat{R}$ within 0.0042 of one (see Section 2 of the Supplementary Materials). 

Figure \ref{fig:seroSpagetti} shows the posterior mean seroprevalence curves for each study location. While most curves do not show age-specific trends, a few show clear deviations, notably, Bogot\'a, Colombia and Chennai, India. The 95\% credible interval of the seroprevalence function did not contain a flat trend for both of these locations. Both showed a decreased seroprevalence for the oldest individuals. While most locations showed roughly flat seroprevalence, the average seroprevalence varies substantially across locations. This is unsurprising as we expect heterogeneity due to factors such as when the seroprevalence study was conducted and when waves of COVID-19 arose in each location.
\begin{figure}[tbh]
    \centering
    \subfigure[]{
    \includegraphics[width=.315\textwidth]{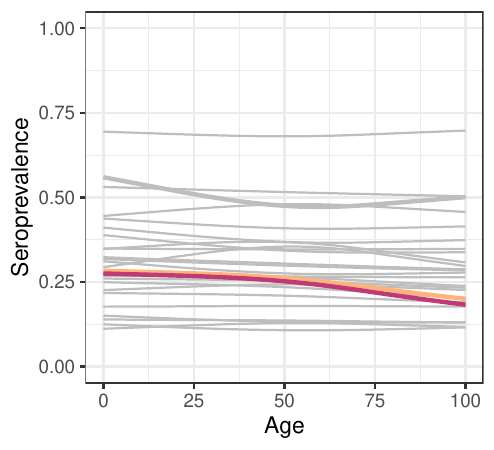}
    }
    \subfigure[]{
    \includegraphics[width=.315\textwidth]{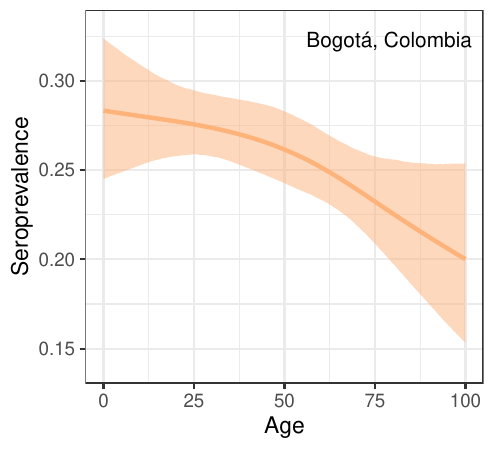}
    }
    \subfigure[]{
    \includegraphics[width=.315\textwidth]{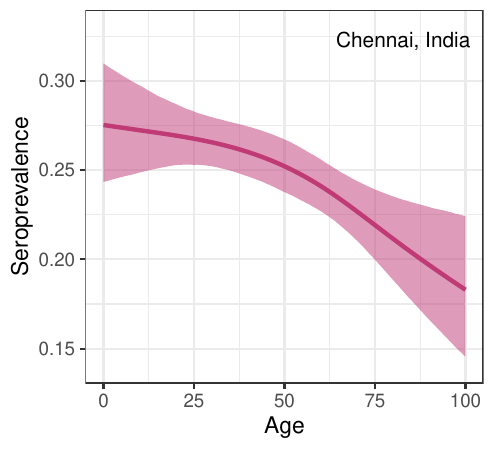}
    }
    \caption{(a) Posterior mean seroprevalence curve for each location, colored to emphasize those locations where seroprevalence varies by age. Panels (b) and (c) show the posterior mean and 95\% credible interval for the locations highlighted in (a). All studies were conducted between June 2020 and March 2021.}
    \label{fig:seroSpagetti}
\end{figure}

\begin{figure}[tb]
    \centering
    \subfigure[]{
        \includegraphics[width=.4\textwidth]{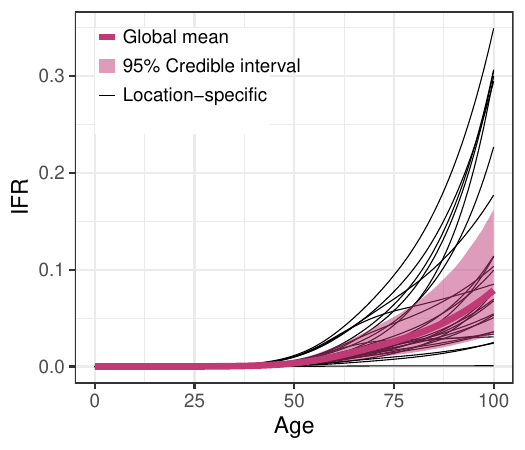}
    }
    \subfigure[]{
        \includegraphics[width=.4\textwidth]{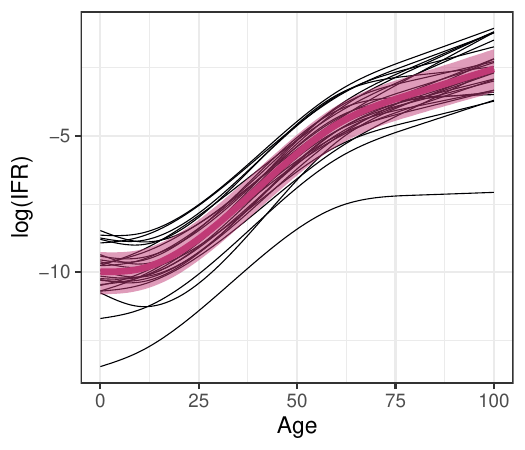}
    }
    \caption{Posterior mean of the global IFR curve with 95\% credible interval for the mean. Panel (a) shows the IFR on its original scale and panel (b) shows the log-scale. The posterior means of location-specific IFR curves are shown in black.}
    \label{fig:globalIFR}
\end{figure}

Figure \ref{fig:globalIFR} shows the global IFR curve as well as the posterior mean IFR curve for each location, where we again use the term ``global" in the modeling rather than geographic sense. Viewing IFR on the log-scale (Figure \ref{fig:globalIFR}(b)), we can see the locations generally follow a s-curve with IFR increasing less quickly or even decreasing with age for children and older individuals. The individual locations show variation in this trend in terms of their intercepts and shapes. IFR posterior mean increased with age for children in some locations but decreased with age in others, although not significantly. Similarly, the IFR flattened out more or less for older individuals depending on the location. These results highlight the need for a model that can estimate location-specific, non-linear IFR curves.

\begin{figure}[h!]
    \centering
    \subfigure[]{
			\includegraphics[width=.45\textwidth]{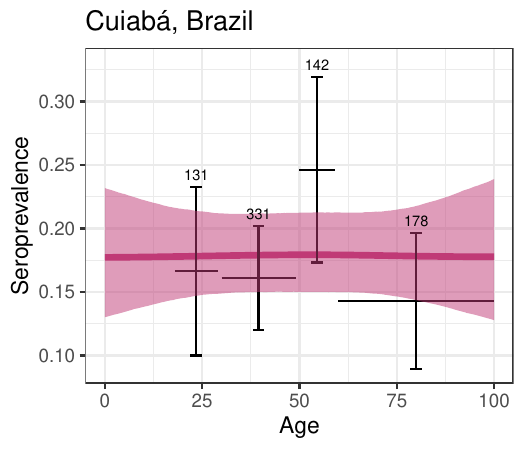}
}
	\subfigure[]{
			\includegraphics[width=.45\textwidth]{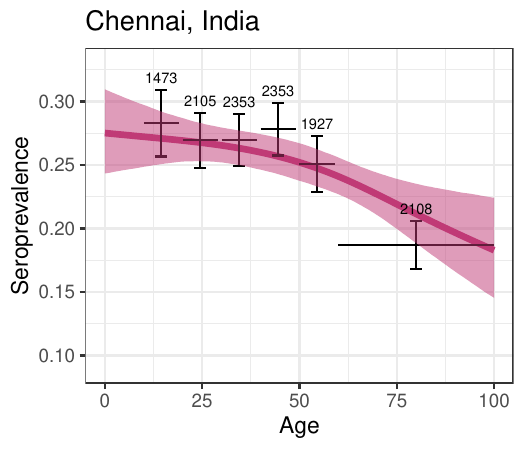}
}
    \subfigure[]{
			\includegraphics[width=.45\textwidth]{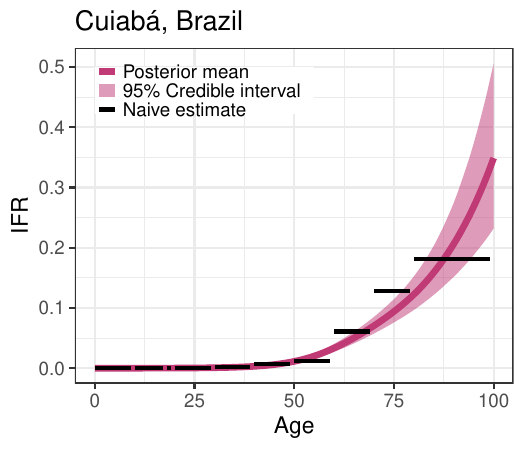}
}
	\subfigure[]{
			\includegraphics[width=.45\textwidth]{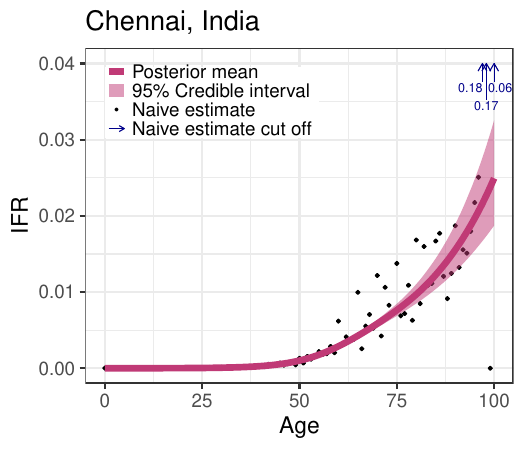}
}
    \caption{Panels (a) and (b) show the posterior mean and 95\% credible interval for (a) Cuiab\'a, Brazil's and (b) Chennai, India's seroprevalence curves, annotated with the seroprevalence study sample size for each bin. The Rogan-Gladen estimators and approximate 95\% confidence intervals are shown as error bars. Panels (c) and (d) show the posterior mean and 95\% credible interval for the IFR curves. Naive estimates for the IFR are shown as points when single year age bins are available for the death data and as black lines when the death data is binned. }
    \label{fig:karnataka}
\end{figure}
Figure \ref{fig:karnataka} focuses on the seroprevalence and IFR functions for Cuiab\'a, Brazil and Chennai, India. The posterior mean and 95\% credible intervals are compared to naive estimates at the age bin level. For seroprevalence, the Rogan-Gladen estimate is shown, which adjusts the raw prevalence by estimates of the test sensitivity and specificity \citep{rogan1978estimating}. For IFR, the naive estimate represents the death rate divided by the Rogan-Gladen seroprevalence estimate, assuming the death rate and seroprevalence are constant within an age bin.

We can see the strength of modeling the seroprevalence and IFR as smooth functions of age rather than modeling each age bin separately. 
We do not expect seroprevalence or IFR to largely jump across consecutive age bins as is observed in the naive estimates, but rather expect some smooth underlying function.
For example, the 50-59 serology age bin in Cuiab\'a had a smaller sample size relative to the surrounding bins and showed an unusual jump in seroprevalence. 
Rather than fitting a seroprevalence curve that goes through the naive seroprevalence estimate, the model pools information across adjacent age bins. 
However, the seroprevalence curve can overcome the mean zero prior when the data suggests it, as is shown for Chennai, India (see panel (b) of Figure \ref{fig:karnataka}). Additionally, by sharing information across age bins, the uncertainty of our seroprevalence function is smaller than the uncertainty of the naive point estimates in most cases. 

Figure \ref{fig:karnataka} again shows the benefits of estimating IFR as a continuous function of age. In Cuiab\'a, the death data is aggregated into less than 10 bins. By estimating a smooth, underlying function of IFR, we are able to estimate IFR for each specific age, not just averages for the age bins the data is available at.  
In Chennai, the IFR function passes through roughly the center of the naive estimates, estimating a smooth function given the extremely noisy single year naive estimates (see Figure \ref{fig:karnataka}(d)). 
Seroprevalence and IFR functions for each of the 26 locations in our analysis are included in Sections 4 and 5 of the Supplementary Materials.
    
\begin{figure}
    \centering
    \includegraphics[width=.99\textwidth, clip, trim=0in .05in 0in .1in]{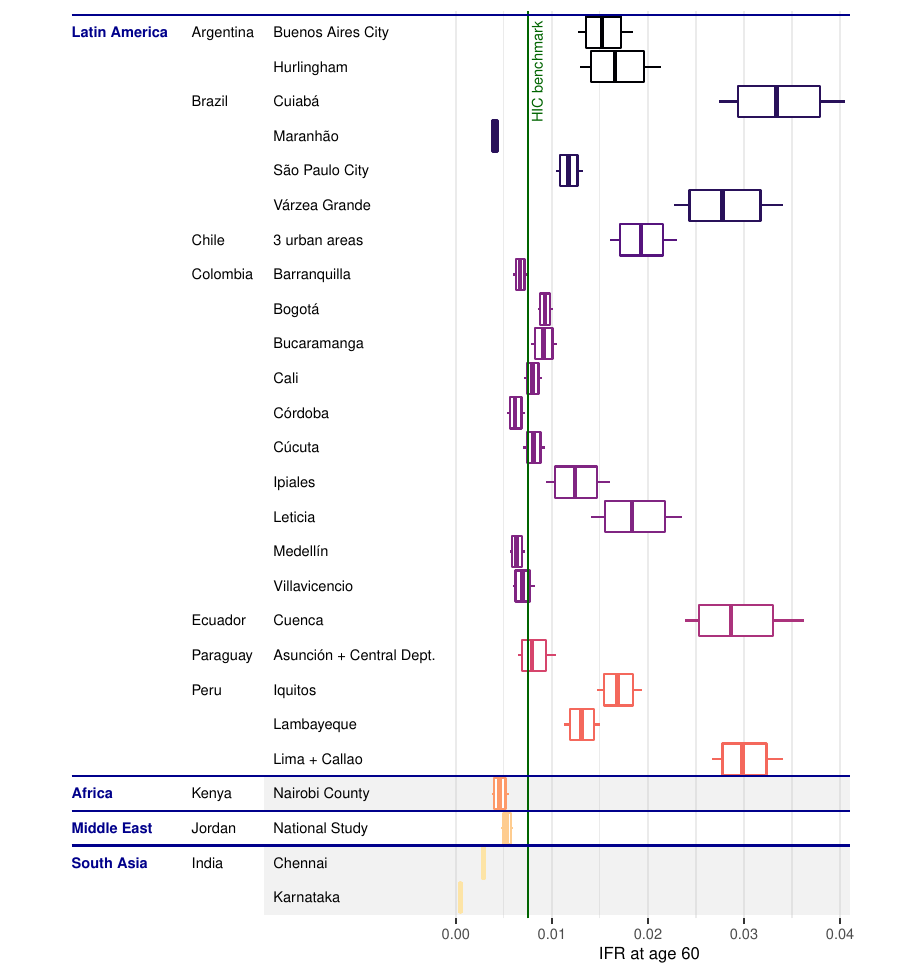}
    \caption{IFR at age 60 for each location. Whiskers indicate 95\% credible intervals, boxes indicate 80\% credible intervals, and the center line indicates the posterior median of the posterior distribution for age-60 IFR. The high-income countries (HIC) benchmark\citep{eje} is shown as a vertical line. Locations with a grey background have less than 50\% of deaths well certified \citep{fullman2017measuring}.}
    \label{fig:age60}
\end{figure}
Figure \ref{fig:age60} provides estimates and credible intervals for IFR at age sixty at all locations. Because we model IFR as a continuous function of age, we can estimate IFR for this specific age rather than being restricted to discussing the average IFR over the age bin at which the data was reported. These location-specific estimates are compared to the estimated IFR of high-income countries at age 60 given in Levin et al\citep{eje}. Note, this high-income country benchmark was deemed an appropriate comparator for the studies in this data set by Levin et al\citep{ours} in terms of the timing and the inclusion criteria of the seroprevalence studies and seroreversion/death reporting adjustments made. We can see most locations have an IFR near or above the high-income countries estimate at age 60. Of the eight study locations below the benchmark, three have less than 50\% of all deaths well certified, meaning less than 50\% of deaths in that country are registered to a specific, well-defined cause \citep{fullman2017measuring}. These three are Nairobi County, Kenya, Chennai, India, and Karnataka, India. Combining this result with the findings in Knutson et al\citep{knutson2022estimating}, which found COVID-19 deaths are likely to be undercounted in many developing countries, suggests the estimated IFR may be below the high-income country benchmark because of an under-reporting of COVID-19 deaths, rather than an actual decrease in the risk of COVID-19 at these locations. 
For example, Silva et al\citep{silva2020population} found the increased number of natural cause deaths for those over the age of 60 during the pandemic compared to before was around two times the number of reported COVID-19 deaths in Maranh\~ao, Brazil. If the increased number of natural cause deaths could be attributed to unreported COVID-19 deaths, the IFR estimate would approximately double and exceed the high-income country benchmark.

If we view the number of reported deaths in our study locations as minimums, with death reports either accurate or an underreport, then our IFR estimates can be interpreted as lower-bounds for the true IFR. In this case, COVID-19 has been catastrophic for some developing country locations like Cuiab\'a, Brazil and Lima, Peru that have an IFR at least three times that of high-income countries for those aged sixty. 

\newpage

\begin{figure}[htb]
    \centering
    \includegraphics[width=\textwidth]{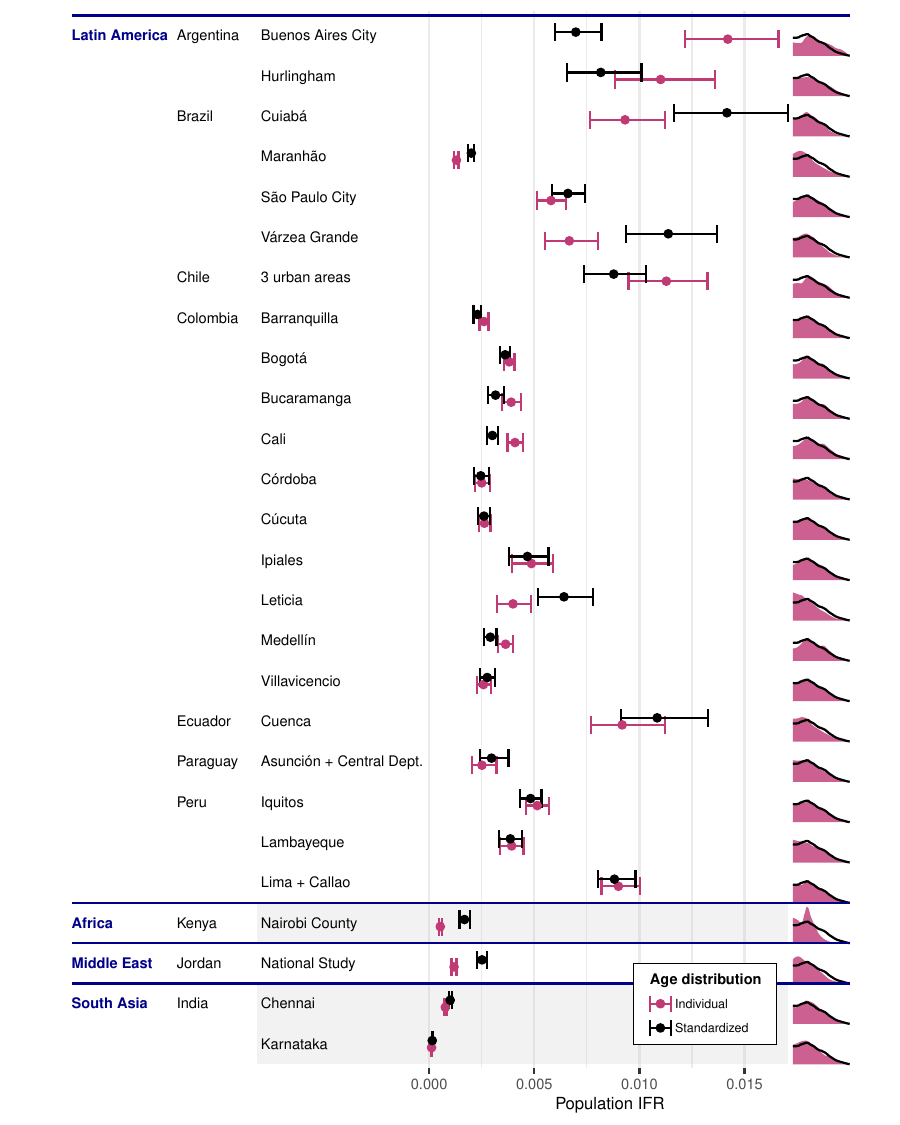}
    \caption{The posterior mean population IFR (the point) with a 95\% credible interval. Estimates are based on the location-specific age distribution (``Individual") or based on the median age distribution across our study locations (``Standardized"). Locations with a grey background have less than 50\% of deaths well certified \citep{fullman2017measuring}. The age distribution for each location (filled) compared to the standardized age distribution (the line) is shown on the right.}
    \label{fig:popIFR}
\end{figure}
Figure \ref{fig:popIFR} shows the effect of the age distribution on the population IFR. In Buenos Aires City, Argentina and Cali, Colombia the population IFR was significantly higher when based on their own age distributions compared to the standardized age distribution. This is because the average age was smaller for the standardized distribution than that for their specific locations. In the case of Buenos Aires City, the population IFR estimate was considerably closer to the other locations' when controlling for the age distribution. 
We see the opposite at locations like Maranh\~ao, Nairobi County, and the Jordan national study, where the average age was smaller for the specific locations than the standardized distribution. The standardized distribution estimates were closer to the other locations for these locations as well.
By estimating seroprevalence and IFR as functions of age, we are able to separate out the effects of the seroprevalence, IFR, and age distribution on the population level estimate.

\section{Discussion} \label{discussion}

In this work, we introduced an adaptable Bayesian hierarchical model that enables estimation of location- and age-specific IFR and seroprevalence curves, which accurately reflect uncertainty due to limited data and uncertainty in the underlying test characteristics. The Bayesian framework presents a natural framework for propagating test characteristic uncertainty and allowed us to incorporate prior knowledge to improve identifiability. 

Apply this model to our developing countries dataset, we found seroprevalence was not dependent on age for most locations, while IFR increased non-linearly with age. The IFR curves were diverse, emphasizing the need for location-specific estimates. IFR for those aged 60 in our study locations was near or above the age 60 high-income country benchmark for most locations. Finally, we showed the importance of considering the location's age distribution by comparing the population IFR using the location-specific age distribution to the population IFR using a standardized age distribution.

While our model was tailored to the developing countries COVID-19 data set, the model is applicable to estimating IFR for any disease using seroprevalence studies and reported deaths. 
Specifically, this methodology may be particularly useful for modeling novel emerging pathogens. During the initial surge of COVID-19, one of the primary issues noted in public communications was the challenge of incorporating uncertainty into estimates (see, e.g., Dean\citep{dean2020}, Foad \citep{foad2021}). Our model may be useful for estimating IFR for future novel pathogens, when it is essential to 
 maintain sufficient model flexibility, such as avoiding assumptions about the relationship of mortality with age, before much is known about the disease.  

Our model is flexible and can meet the unique challenges and opportunities of future applications. 
For example, additional covariates beyond age could be added to the vector of covariates for serology, $\vec{z}_{\ell,a}$, or deaths, $\vec{x}_{\ell,a}$. While we chose not to pool information across the seroprevalence coefficients in our application,  
information could be pooled following the hierarchical modeling framework presented for the IFR coefficients. 

\bibliographystyle{WileyNJD-AMA.bst}
\bibliography{mybib}

\begin{thebibliography}{10}
\providecommand \doibase [0]{http://dx.doi.org/}%

\bibitem{jha2022counting}
Jha P, Brown PE, Ansumana R. Counting the global {COVID-19} dead. {\it The
  Lancet} 2022\string; 399(10339)\string: 1937--1938.

\bibitem{deaths}
{Our World in Data} . {COVID}-19 Data Explorer.
  {https://ourworldindata.org/explorers/coronavirus-data-explorer};  2022.
\newblock Accessed: 2022-11-13.

\bibitem{eje}
Levin AT, Hanage WP, Owusu-Boaitey N, Cochran KB, Walsh SP, Meyerowitz-Katz G.
  Assessing the age specificity of infection fatality rates for {COVID}-19:
  {Systematic} review, meta-analysis, and public policy implications. {\it
  European Journal of Epidemiology} 2020\string; 35(12)\string: 1123--1138.

\bibitem{pezzullo2023age}
Pezzullo AM, Axfors C, Contopoulos-Ioannidis DG, Apostolatos A, Ioannidis JP.
  Age-stratified infection fatality rate of {COVID-19} in the non-elderly
  population. {\it Environmental Research} 2023\string; 216\string: 114655.

\bibitem{starke2021isolated}
Starke KR, Reissig D, Petereit-Haack G, Schmauder S, Nienhaus A, Seidler A. The
  isolated effect of age on the risk of {COVID-19} severe outcomes: {A}
  systematic review with meta-analysis. {\it BMJ Global Health} 2021\string;
  6(12)\string: e006434.

\bibitem{national2020evaluating}
{National Academies of Sciences, Engineering, and Medicine} . {\it Evaluating
  data types: {A} guide for decision makers using data to understand the extent
  and spread of {COVID-19}}.
\newblock Washington, DC: The National Academies Press .
\newblock 2020

\bibitem{gelman2020bayesian}
Gelman A, Carpenter B. {Bayesian} analysis of tests with unknown specificity
  and sensitivity. {\it Journal of the Royal Statistical Society: Series C
  (Applied Statistics)} 2020\string; 69(5)\string: 1269--1283.

\bibitem{malani2022vaccine}
Malani A, Soman S, Ramachandran S, Chen A, Lakdawalla DN. Vaccine Allocation
  Priorities Using Disease Surveillance and Economic Data. tech. rep., National
  Bureau of Economic Research; Massachusetts, USA:   2022

\bibitem{perez}
Perez-Saez J, Lauer SA, Kaiser L, et al. Serology-informed estimates of
  {SARS-CoV-2} infection fatality risk in {Geneva}, {Switzerland}. {\it The
  Lancet Infectious Diseases} 2021\string; 21(4)\string: e69--e70.

\bibitem{odriscoll}
O’Driscoll M, Ribeiro Dos~Santos G, Wang L, et al. Age-specific mortality and
  immunity patterns of {SARS-CoV-2}. {\it Nature} 2021\string;
  590(7844)\string: 140--145.

\bibitem{ours}
Levin AT, Owusu-Boaitey N, Pugh S, et al. Assessing the burden of {COVID-19} in
  developing countries: {Systematic} review, meta-analysis and public policy
  implications. {\it BMJ Global Health} 2022\string; 7(5)\string: e008477.

\bibitem{lancet}
{COVID-19 Forecasting Team} . Variation in the {COVID-19} infection-fatality
  ratio by age, time, and geography during the pre-vaccine era: {A} systematic
  analysis. {\it The Lancet} 2022\string; 399(10334)\string: 1469--1488.

\bibitem{ourdata}
Levin A, Meyerowitz-Katz G, Soman S, et al. covid-ifr-data v1.0.1.
  http://doi.org/10.5281/zenodo.8048531;  2023

\bibitem{stringhini2020seroprevalence}
Stringhini S, Wisniak A, Piumatti G, et al. Seroprevalence of {anti-SARS-CoV-2}
  {IgG} antibodies in {Geneva, Switzerland} ({SEROCoV-POP}): {A}
  population-based study. {\it The Lancet} 2020\string; 396(10247)\string:
  313--319.

\bibitem{larremore2022optimizing}
Larremore DB, Fosdick BK, Zhang S, Grad YH. Optimizing prevalence estimates for
  a novel pathogen by reducing uncertainty in test characteristics. {\it
  Epidemics} 2022\string; 41\string: 100634.

\bibitem{jacofsky2020understanding}
Jacofsky D, Jacofsky EM, Jacofsky M. Understanding antibody testing for
  {COVID-19}. {\it The Journal of Arthroplasty} 2020\string; 35(7)\string:
  S74--S81.

\bibitem{karanikolos2020comparable}
Karanikolos M, McKee M, others . How comparable is {COVID-19} mortality across
  countries?. {\it Eurohealth} 2020\string; 26(2)\string: 45--50.

\bibitem{anup}
Cai R, Novosad P, Tandel V, Asher S, Malani A. Representative estimates of
  {COVID-19} infection fatality rates from four locations in {India}:
  {Cross}-sectional study. {\it BMJ {Open}} 2021\string; 11(10)\string:
  e050920.

\bibitem{cdcguidelines}
{Centers for Disease Control and Prevention} . Interim Guidelines for
  {COVID}-19 Antibody Testing.
  {https://www.cdc.gov/coronavirus/2019-ncov/lab/resources/antibody-tests-guidelines.html};
  2021.
\newblock Accessed: 2021-01-07.

\bibitem{esteve2020national}
Esteve A, Permanyer I, Boertien D, Vaupel JW. National age and coresidence
  patterns shape {COVID-19} vulnerability. {\it Proceedings of the National
  Academy of Sciences} 2020\string; 117(28)\string: 16118--16120.

\bibitem{ruggles2008intergenerational}
Ruggles S, Heggeness M. Intergenerational coresidence in developing countries.
  {\it Population and Development Review} 2008\string; 34(2)\string: 253--281.

\bibitem{stan}
{Stan Development Team} . {RStan}: {The} {R} interface to {Stan}.
  https://mc-stan.org/;  2022.
\newblock R package version 2.21.5.

\bibitem{gelman1995bayesian}
Gelman A, Carlin JB, Stern HS, Rubin DB. {\it Bayesian Data Analysis}.
\newblock Chapman and Hall/CRC .
\newblock 1995.

\bibitem{rogan1978estimating}
Rogan WJ, Gladen B. Estimating prevalence from the results of a screening test.
  {\it American Journal of Epidemiology} 1978\string; 107(1)\string: 71--76.

\bibitem{fullman2017measuring}
Fullman N, Barber RM, Abajobir AA, et al. Measuring progress and projecting
  attainment on the basis of past trends of the health-related {Sustainable
  Development Goals} in 188 countries: {An} analysis from the {Global Burden of
  Disease Study} 2016. {\it The Lancet} 2017\string; 390(10100)\string:
  1423--1459.

\bibitem{knutson2022estimating}
Knutson V, Aleshin-Guendel S, Karlinsky A, Msemburi W, Wakefield J. Estimating
  Global and Country-Specific Excess Mortality During the {COVID-19} Pandemic.
  {\it arXiv preprint arXiv:2205.09081} 2022.

\bibitem{silva2020population}
Silva dAAM, Lima-Neto LG, Azevedo d.~M. P. e. S.~dC, et al. Population-based
  seroprevalence of {SARS-CoV-2} and the herd immunity threshold in
  {Maranh{\~a}o}. {\it Revista de Saude Publica} 2020\string; 54.

\bibitem{dean2020}
Dean J. Experts: {A}cknowledge uncertainty in {COVID} communication.
  {https://news.cornell.edu/stories/2020/09/experts-acknowledge-uncertainty-covid-communication};
  2020.
\newblock Accessed: 2023-1-04.

\bibitem{foad2021}
Foad C. Embracing pandemic uncertainty in science, society and policy.
  {https://royalsociety.org/blog/2021/07/embracing-pandemic-uncertainty-in-science-society-and-policy/};
  2021.
\newblock Accessed: 2023-1-04.

\end{thebibliography}

\newpage

\appendix

\section{Approximating the age distribution} \label{definePop}
The exact age distribution is not available for any particular location. Instead, we have the age distribution split into age bins that do not necessarily correspond to those of the serology study or death data. Let $\mathcal{A}$ be the set of such disjoint age bins corresponding to the age distribution data. Let $\hat{f_\ell}(A)$ denote the the observed proportion of the population in age bin $A$, for $A \in \mathcal{A}$. Then we define the step function 
\begin{align} \label{expandF}
    \tilde{f}_\ell(a) &= \sum_{A \in \mathcal{A}} \frac{\hat{f_\ell}(A)}{|A|} \Ind (a \in A)
\end{align}
where $|A|$ gives the length of interval $A$. 

When the age bins for a population were larger than five, we further refined the step function using the national age distribution for that location, $\hat{f}_{n_\ell}(A)$, which was expanded using \eqref{expandF} to $\tilde{f}_{n_\ell}(a)$. We then define
\begin{align}\label{eq:nat}
    \tilde{f}_\ell(a) &= 
    \sum_{A \in \mathcal{A}} \hat{f_\ell}(A)  
    \frac{\tilde{f}_{n_\ell}(a)}{\sum_{b \in A \cap \mathbb{N}} \tilde{f}_{n_\ell}(b)} 
    \Ind (a \in A)
\end{align}
where $\mathbb{N}$ denotes the natural numbers. Thus, the national age distribution is rescaled so the proportion of the population in $A$ equals $\hat{f}_\ell (A)$, the proportion for location $\ell$. 

\begin{figure}[h!]
    \centering
    \includegraphics[scale=.8]{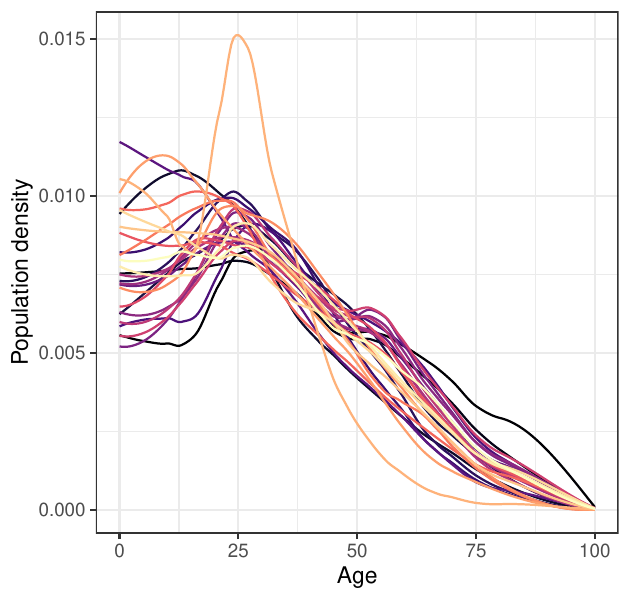}
    \caption{Population density for each location, $f_\ell(a)$.}
    \label{fig:popdens}
\end{figure}

Finally, we used locally weighted smoothing (LOESS) to smooth the population age distribution, rather than using the step function defined in \eqref{eq:nat}. We used the points from the step function up to age 85, as this was the limit of our data in some locations. We added a data point at age 100 with a value of 0 as individuals over the age of 100 are very rare in developing countries. If any predictions were less than zero, we subtracted the minimum values from all the predictions to give all predictions a positive value. This made minimal changes as the negative values were always negligible. Finally, we rescaled the predictions to sum to one. These resulting predictions, shown in Figure \ref{fig:popdens}, are what we defined to be $f_\ell(a)$.

\newpage

\section{Convergence diagnostics}

\FloatBarrier

\begin{figure}[!htb]
    \centering
    \subfigure[]{
			\includegraphics[width=.293\textwidth]{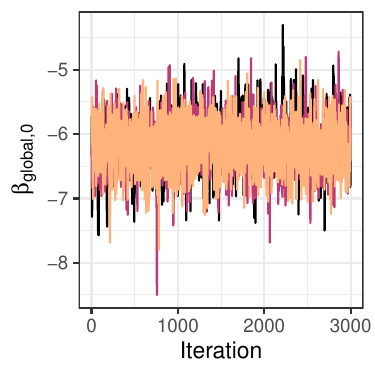}
} \subfigure[]{
			\includegraphics[width=.293\textwidth]{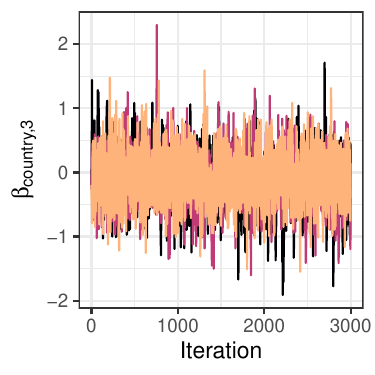}
} \subfigure[]{
			\includegraphics[width=.36\textwidth]{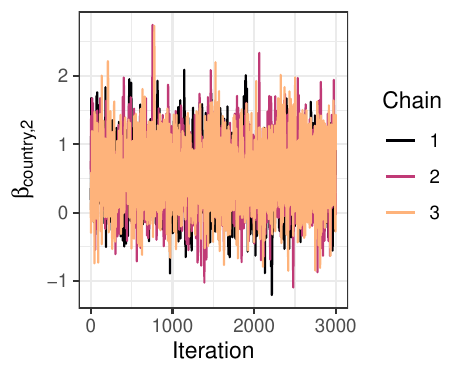}
}
    \caption{Traceplots of (a) $\beta_{\text{global},0}$, (b) $\beta_{\text{country},3}$, and (c) $\beta_{\text{country},2}$, colored by chain.}
    \label{fig:trace}
\end{figure}

 The range of  $\hat{R}$ and effective sample size are shown for groupings of parameters in Figure \ref{tab:convergence} \citep{gelman1995bayesian}. All effective samples sizes were above 1000 and $\hat{R}$ were within 0.0042 of one for each parameter. The traceplots for the three parameters with the smallest effective samples sizes are shown in Figure \ref{fig:trace}. Traceplots for the other parameters are similar, suggesting convergence. Sampling via RStan took 129 minutes using a standard laptop with a 2.7 GHz quad-core processor and 16 GB of RAM.

\begin{figure}[h]
    \centering
    \begin{tabular}{rrr}
      \hline
     & $\hat{R}$ & ESS \\ 
      \hline
   $\beta_{\ell,i}$&$(0.9998, 1.0014)$&$(5673, 11765)$\\ 
$\beta _{\text{global},i}$&$(1.0002, 1.0042)$&$(1338, 3634)$\\ 
$\beta_{\text{country},c_\ell}$&$(1.0007, 1.0037)$&$(1419, 2493)$\\ 
$\sigma_i$&$(1.0001, 1.0002)$&$(2936, 3517)$\\ 
$\sigma_\text{country}$&$(1.0005, 1.0005)$&$(2952, 2952)$\\ 
$\gamma_{\ell,j}$&$(0.9998, 1.0017)$&$(5109, 15224)$\\ 
$\sens_t$&$(1.0000, 1.0014)$&$(2929, 11443)$\\ 
$\spec_t$&$(0.9998, 1.0006)$&$(5869, 14531)$\\ 
      \hline
    \end{tabular}
    \caption{\label{tab:convergence} The range of $\hat{R}$ and the effective sample size (ESS) for each grouping of parameters.}
\end{figure}

\FloatBarrier

\newpage

\section{Seroprevalence curves by study date}

\begin{figure}[!ht]
    \centering
    \includegraphics[scale=1]{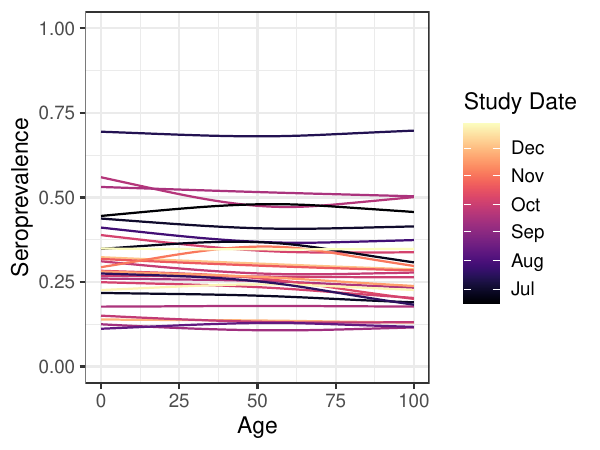}
    \caption{Posterior mean seroprevalence curve for each study location, colored by the start date of each study. }
    \label{fig:seroDate}
\end{figure}

\FloatBarrier

\newpage

\section{Seroprevalence curves for each location}

Each of the following plots shows the posterior mean seroprevalence curve with a 95\% credible interval. Also shown is the Rogan-Gladen estimate with an approximate confidence interval that treats sensitivity and specificity as known. 
\begin{center}
\includegraphics[width=.95\textwidth, page=1, clip, trim=0in 0in 0in 3in]{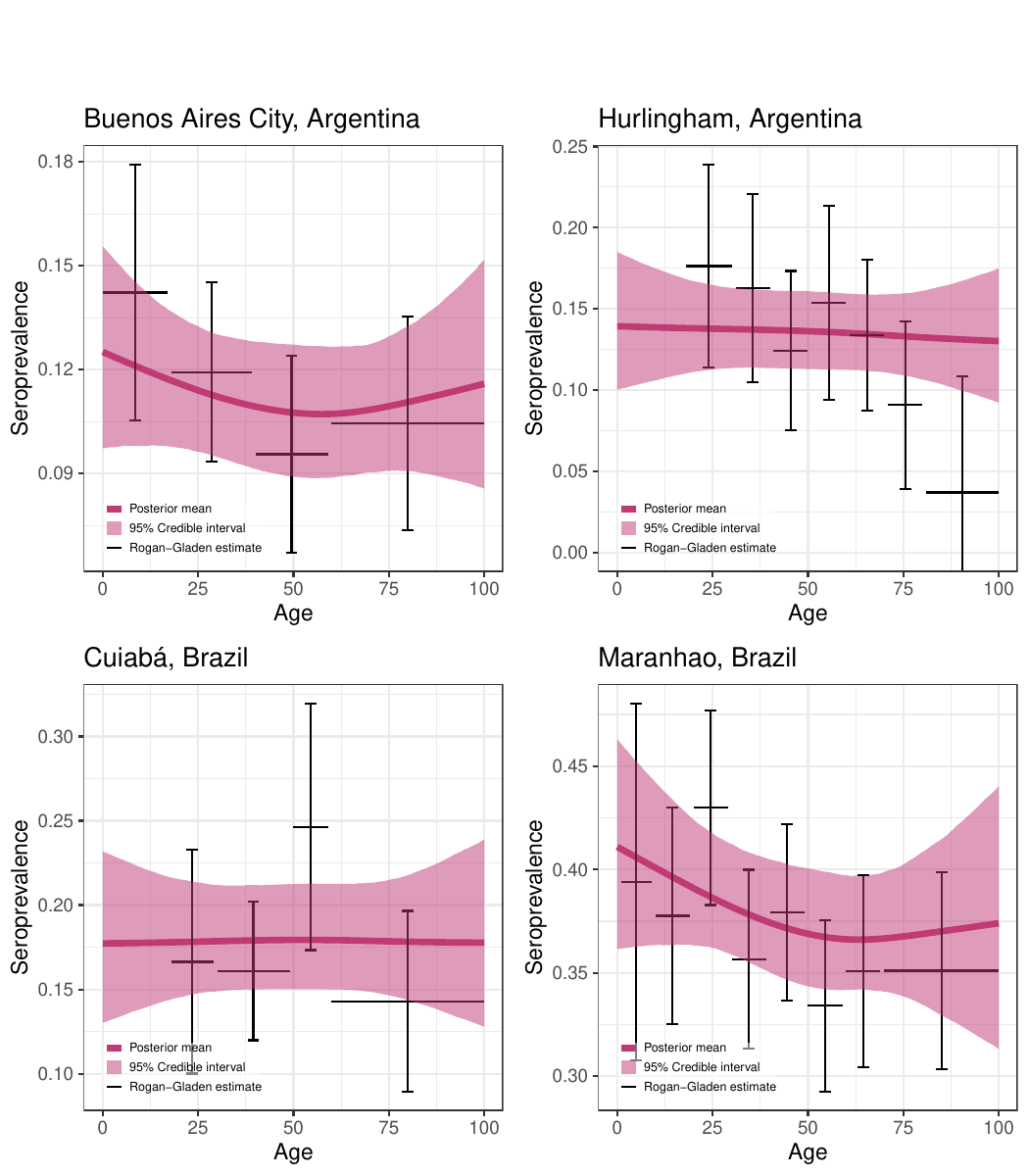}
\includegraphics[width=.95\textwidth, page=2]{Plots/6-30-final-run_allSeroV2.pdf}
\includegraphics[width=.95\textwidth, page=3]{Plots/6-30-final-run_allSeroV2.pdf}
\includegraphics[width=.95\textwidth, page=4]{Plots/6-30-final-run_allSeroV2.pdf}
\includegraphics[width=.95\textwidth, page=5, clip, trim=0in 3in 0in 0in]{Plots/6-30-final-run_allSeroV2.pdf}
\end{center}

\newpage

\section{IFR curves for each location}

Each of the following plots shows the posterior mean IFR curve with a 95\% credible interval.  Naive estimates for the IFR are calculated as the empirical death rate divided by the Rogan-Gladen estimator for seroprevalence. Note, when data is only available at the bin level, death and seroprevalence rates are assumed uniform within the age bin for the naive estimate. 

\begin{center}
\includegraphics[width=.95\textwidth, page=1, clip, trim=0in 0in 0in 3in]{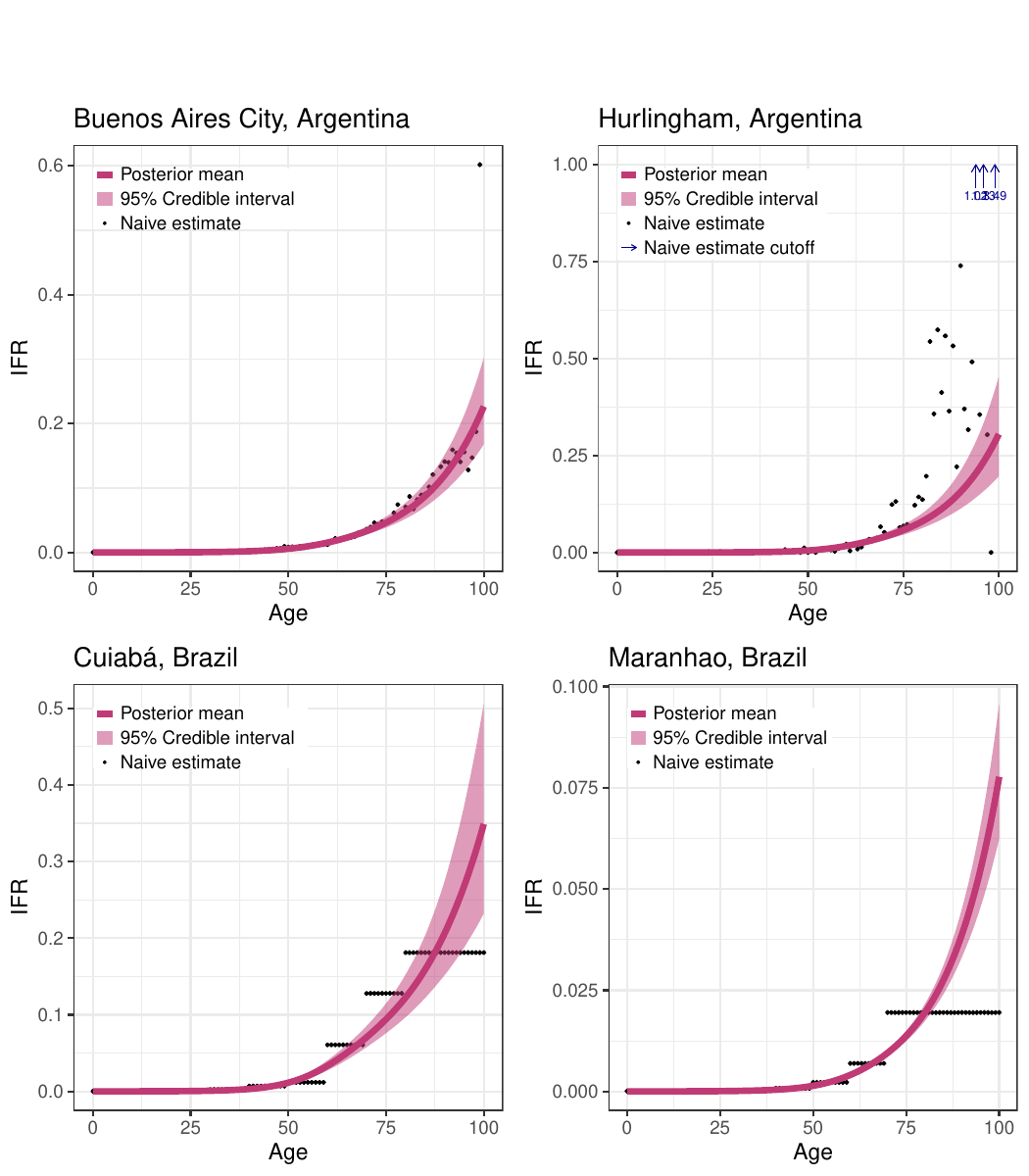}
\includegraphics[width=.95\textwidth, page=2]{Plots/6-30-final-run_allIFRV2.pdf}
\includegraphics[width=.95\textwidth, page=3]{Plots/6-30-final-run_allIFRV2.pdf}
\includegraphics[width=.95\textwidth, page=4]{Plots/6-30-final-run_allIFRV2.pdf}
\includegraphics[width=.95\textwidth, page=5, clip, trim=0in 3in 0in 0in]{Plots/6-30-final-run_allIFRV2.pdf}
\end{center}

\end{document}